\documentclass[]{aa}
\usepackage{graphicx}

\include{psfig}

\begin{document}

\title{Atmospheric escape from hot Jupiters}

\author{
A.~Lecavelier des Etangs \inst{1} 
\and
A.~Vidal-Madjar  \inst{1} 
\and
J.~C.~McConnell \inst{2} 
\and
G.~H\'ebrard \inst{1} 
}

\offprints{A. Lecavelier des Etangs,
   \email{lecaveli@iap.fr}}

\institute{Institut d'Astrophysique de Paris, CNRS, 98 bis boulevard~Arago, 
F-75014 Paris, France
         \and
        Department of Earth and Atmospheric Science, 
  York University, North York, Ontario, Canada
}

\date{Received ...; accepted ...}

\abstract{
The extra-solar planet 
HD\,209458b has been found to have an
extended atmosphere of escaping atomic hydrogen (Vidal-Madjar et al. 2003),
suggesting that ``hot Jupiters'' closer to their parent stars could evaporate.
Here we estimate the atmospheric escape (so called evaporation rate) 
from hot Jupiters and their corresponding life time against evaporation.
The calculated evaporation rate of HD\,209458b is in excellent agreement with 
the H\,{\sc i} Lyman-$\alpha$ observations.
We find that the tidal forces and high temperatures in the upper 
atmosphere must be taken into account 
to obtain reliable estimate of the atmospheric escape. 
Because of 
the tidal forces, we show that there is a new escape mechanism at
intermediate temperatures at which the exobase reaches the Roche lobe.
From an energy balance, we can estimate plausible values for the planetary
exospheric temperatures, and thus obtain typical life times of
planets as a function of their mass and orbital distance.


}

   \maketitle
%

\section{Introduction}
Among the more than one hundred extra-solar planets known,
over 15\% are closer than 0.1\,AU from the central star.
Except for a possible transiting planet (OGLE-TR-56b, Konacki et al. 2003; 
Torres et al. 2003),
there are no detection of planets closer than about 0.04\,AU.
Following the discovery that the planet HD\,209458b shows a
surprisingly large escape of atomic hydrogen (Vidal-Madjar et al.~2003;
confirmed by Vidal-Madjar et al.~2004),
the escape flux from 
hot Jupiters needs to be 
evaluated. Already in 1995, together with the release of the first hot Jupiter 
discovery (Mayor \& Queloz 1995), Burrows \&
Lunine (1995) highlighted the evaporation issue. But 
Guillot et al.~(1996) concluded that the mass loss was not significant. 
However their estimates did not consider two factors
which need to be addressed: 
the influence of the strong tidal forces 
from the neighboring parent stars, 
and the high upper atmosphere temperature
(H\'ebrard et al. 2003).

\section{The upper atmosphere model}

Radiative equilibrium is commonly
used to calculate the temperature of the upper atmosphere 
(see Schneider et al.\ 1998). 
But this is not appropriate because it does not apply to the
low density upper atmosphere. 
As an example, 
in the Solar system, 
the temperature of planetary upper atmospheres (thermosphere, exosphere)
is much higher than $T_{\rm eff}$,
the effective temperature of the lower atmosphere.
For Earth and Jupiter, $T_{\rm eff}$ are $\sim$250\,K and 150\,K, 
while thermospheric temperatures are $\sim$1\,000\,K for both planets
(Chamberlain \& Hunten 1987).
Although the observed high temperatures in the giant planets 
are not yet explained, extreme and far ultraviolet 
fluxes, Solar wind and perhaps gravity waves may contribute 
to the heating in uncertain amounts ({\it e.g.}, Hunten \& Dessler 1977).

\begin{figure}[bthp]
\includegraphics[width=\columnwidth]{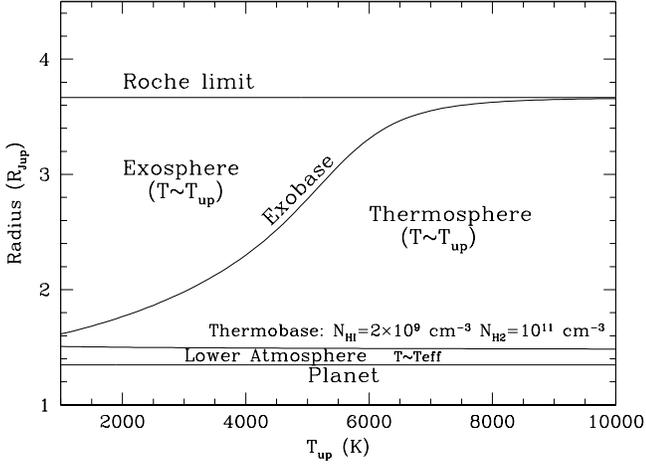}
\caption[]{Vertical structure of the atmosphere of HD\,259458b 
as a function of the temperature of the upper atmosphere
(Model~A).
At the top of the thermosphere, the exobase is the
critical level where the mean free path is equal to the distance
to the Roche lobe. 
For high temperatures, the exobase reaches the Roche lobe;
this leads to a {\it geometrical blow-off}.
}
\label{plot_Rz_vs_T}
\end{figure}

The atmospheric escape flux strongly depends on the temperature
profile of the atmosphere. As the temperature of the upper atmosphere of
extra-solar planets is
not known, in a first step, we will use it as an input parameter: 
$T_{\rm up}$. We consider a simple atmospheric structure similar to that
observed in the giant planets of the Solar system, that is a two
level temperature structure: $T_{\rm eff}$ in the lower atmosphere
and $T_{\rm up}$ above. We assume that the lower atmosphere 
(up to the thermobase) is mainly composed of molecular hydrogen 
at $T_{\rm eff}$. Above that level, 
in the thermosphere and exosphere, the gas is a mixture of atomic 
and molecular hydrogen at $T_{\rm up}$ (Fig.~\ref{plot_Rz_vs_T}). 

The quantitative results presented in this letter do not depend on 
the estimated $T_{\rm eff}$, but on the density at the 
thermobase, $n_{\rm H\,I}$ and $n_{{\rm H}_2}$ for atomic
and molecular hydrogen, respectively. In the following we will use three
different estimates of these densities. In the case of Jupiter and Saturn,
H$_2$ is the main constituent at the thermobase, 
where the temperature transition is observed to occur for
$n_{\rm H\,I}  \simeq 2\times10^9$~cm$^{-3}$ and 
$n_{{\rm H}_2} \simeq 1\times10^{11}$~cm$^{-3}$ (Chamberlain \& Hunten 1987); 
these values will be used for the Model A. By
evaluating the production of atomic hydrogen with realistic condition for
the lower atmosphere of HD\,209458b, Liang et al. (2003) have shown
that the H\,{\sc i} mixing ratio could be about 10\% at the
thermobase. We thus use the corresponding 
$n_{\rm H\,I} = 1\times10^{10}$~cm$^{-3}$ and 
$n_{{\rm H}_2} = 9\times10^{10}$~cm$^{-3}$ for the
intermediate Model B. Finally, at high thermospheric temperatures, it
is likely that most of the molecular hydrogen is dissociated into atomic
hydrogen (Coustenis et al.\ 1998); we use 
$n_{\rm H\,I} =2\times10^{11}$~cm$^{-3}$ and 
$n_{{\rm H}_2}= 0$~cm$^{-3}$ for the corresponding Model C. 

\section{The atmospheric escape}

The atmospheric escape critically depends on atmospheric temperatures.
Below a critical temperature, T$_{\rm c}$, the escape flux
can be calculated using the Jeans escape estimate which refers to the
escape of particles whose velocity is in the tail of the
Boltzmann distribution and that have enough energy to escape the planet's
gravity. For temperatures above T$_{\rm c}$, the kinetic energy of the atoms or
molecules is sufficient to overcome gravitational forces and they can
stream, or blow-off. 
For Jeans escape the flux is calculated at a
critical level corresponding to the exobase (Hunten et al.\ 1989). 
This critical level is the level above which particles
can freely escape without collisions. The exobase is usually defined by
the place above which the mean free path ($1/nQ$) is larger than $H$, 
the scale height of the atmosphere,
where $n$ is the volume density. 
This definition results from the approximate calculation of the integrated 
collisional cross-section ($Q$) from the exobase to infinity: 
$\int_{exobase}^{\infty} n(r)Q dr\approx n_{\rm exo}HQ =1$, 
where $n_{\rm exo}$ is the density at the exobase.  
Here this idea must be generalized to take into account the particular 
geometry of hot Jupiters; we define the exobase by
the place above which the mean free path is larger than the distance to
the Roche lobe of the planet:  $\int_{exobase}^{Roche\,lobe} n(r)Qdr =1$.  
The Roche lobe is the last equipotential around the planet beyond which the
equipotentials are open to infinity or to encompass the star. 
Thus the exobase is the level above which atoms and molecules can 
definitively escape the planet. Note that in the case of an isolated planet, 
because the Roche lobe is at infinite distance, the two definitions are
identical. 
Then, the total escape flux from the planet is given by 
$\dot{M} = 4 \pi \, r_{\rm exo}^2 \times n_{\rm exo} 
\mu v_T / (2 \sqrt{\pi})$ $\times 
e^{-\lambda} (\lambda + 1)$, where $\mu$ is the mass of the escaping
elements (H\,{\sc i} and H$_2$), $r_{\rm exo}$ is the radius of the exobase,
and $v_T$ is the
thermal velocity at temperature $T$ (Chamberlain \& Hunten 1987).  
$\lambda$ is defined by $\lambda = - \chi \, \mu /kT$, where $\chi$ is the
gravitational potential. 
We calculated the density profile in the atmosphere 
using the barometric law: $ \vec{\nabla} n = n \vec{\nabla} \lambda $.
Recapture of escaped particles is not possible because 
of the high radiation pressure pushing away hydrogen atoms at hundred
of kilometer per seconds (Vidal-Madjar et al.\ 2003).

We can see from Fig.~\ref{plot_Rz_vs_T} that between 
the classical Jeans escape at low temperatures and the dynamical blow-off of the 
atmosphere at high temperatures, the tidal forces lead to a new escape 
mechanism at intermediate temperatures.
Indeed, for HD\,209458b ($d$=0.047\,AU, $M_p$=0.69\,M$_{\rm Jup}$, and $R_p$=1.35\,R$_{\rm Jup}$)
and $T_{\rm up}$ above $\sim$5000\,K, 
the exobase comes close to the Roche lobe, and 
the kinetic energy is similar to the potential energy
needed to reach this limit. 
In that case, the escape mechanism is 
a {\it geometrical blow-off} 
which is 
due to the filling up of the Roche lobe 
by the thermosphere pulled up by the tidal forces.
Indeed, the dynamical blow-off is obtained when the escape velocity 
$v_{\rm esc}$ 
is of the order of (or smaller than) the thermal velocity $v_T$,
that is for small value 
of $\lambda_{\rm exo}$ ($\lambda$=$v_{\rm esc}^2/v_T^2$). 
With the characteristics of HD\,209458b, the dynamical blow-off 
takes place for temperatures above 20\,000~K.
However, at temperatures between 5\,000~K and 20\,000~K, although 
$\lambda_{\rm exo}\gg$1, we have
$\lambda_{\rm exo}$$-$$\lambda_{\rm Roche}\le$1, and most
of the gas at the exobase can freely reach the Roche lobe.
This geometrical blow-off is due to the spatial proximity of the
Roche lobe through which the gas can escape the planet.

\section{Tidal forces}

\begin{figure}[htbp]
\includegraphics[height=8.0cm]{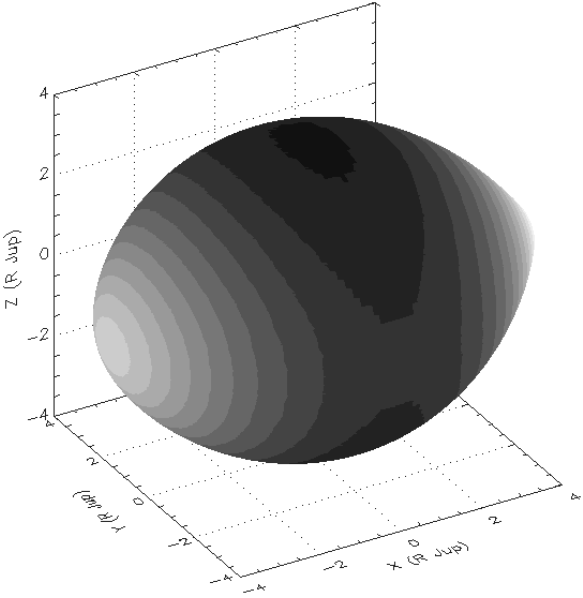}
\caption[]{
Shape of the exobase and 
the corresponding escape rate (Model A, $T_{\rm up}$=11100\,K). 
Light grey is for the large escape rates 
($10^{10}$\,g\,s$^{-1}$\,str$^{-1}$) and black is 
for the small escape rates
($<$5$\times10^9$\,g\,s$^{-1}$\,str$^{-1}$).
The star is on the X-axis toward positive coordinates. The
escape rate is the largest toward the star and in the opposite direction.}
\label{lobe_3d}
\end{figure}

\begin{figure}[hbpt]
\includegraphics[width=\columnwidth]{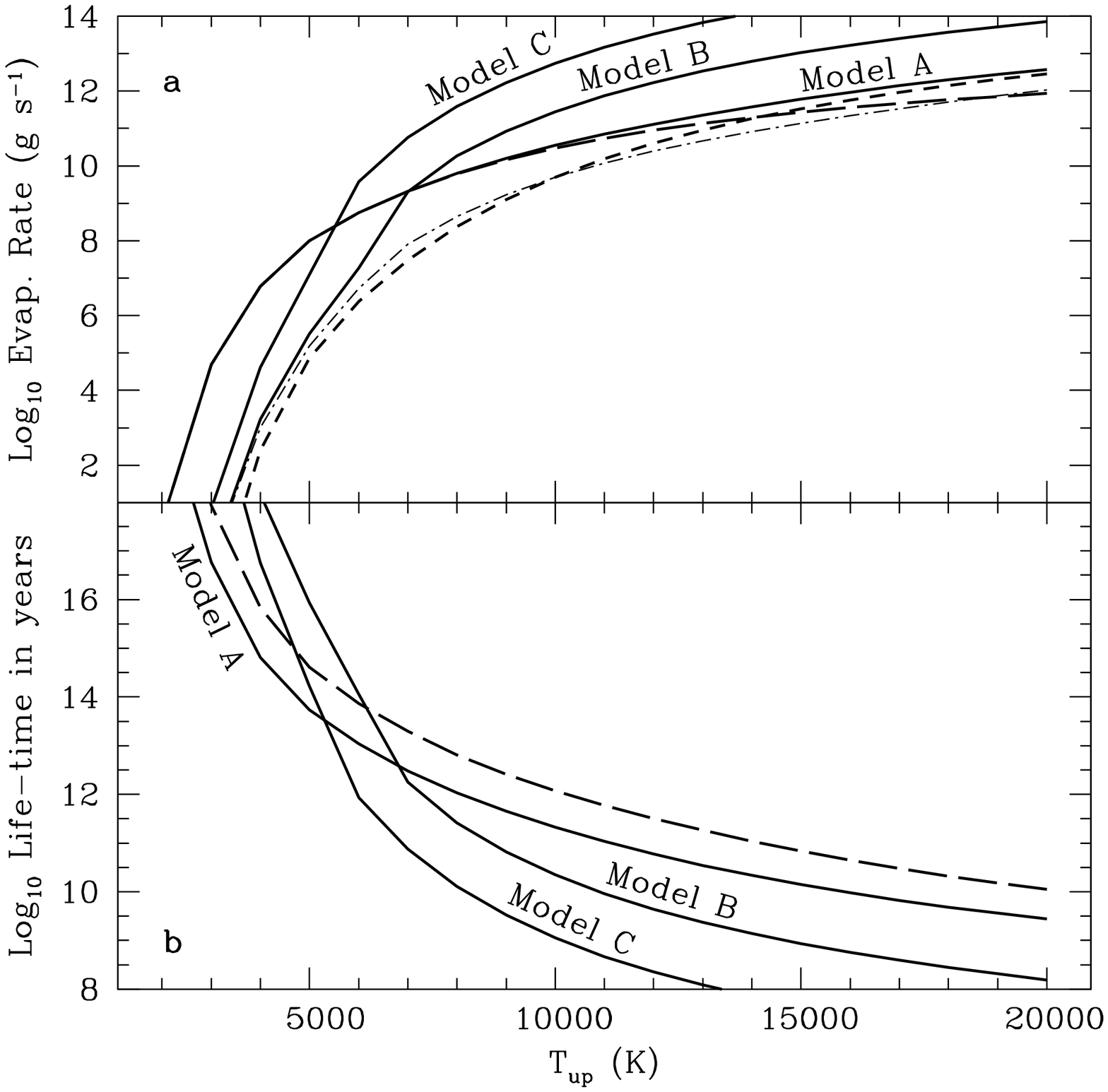}
\caption[]{Escape flux and life time of HD209458b 
as a function of $T_{\rm up}$.
The results from various models show the 
uncertainty due to the lack of constraint on the densities at
the thermobase.

({\bf a}) shows the escape flux of H\,{\sc i} (long dashed) and H$_2$
(short dashed) in the Model A, 
and the total escape flux (solid lines) for models A, B and C. For the Model A, 
the total escape flux from an isolated planet 
(without tidal forces) is plotted with the dot-dashed line. 
The tidal forces increase 
the evaporation rate
by one to two orders of magnitude.

({\bf b}) shows the planet life time. 
A first estimate is obtained from the instantaneous escape flux 
for Model A ($t_1$, dashed line). 
Taking into account the planet's mass
decrease with time, we find shorter life times 
($t_2$, solid lines).
}
\label{figure2}
\end{figure}

In the case of hot Jupiters, the planet's gravity is substantially
modified by stellar tidal forces. 
The common assumption has been to neglect tidal forces by considering
isolated planets far from their star. But tidal forces have a significant
influence on the density distribution in the upper atmosphere of hot
Jupiters.  
In order to calculate the potential energy $\chi$, we include the difference
between the stellar gravitation and the centrifugal effect in the orbiting
planet reference frame, which results in the tidal forces. The
equipotentials are modified, from quasi-spherical in the lower
atmosphere to asymmetric elongated shapes at the level of the
Roche lobe. As a result, 
$\chi$, 
hence the vertical density distribution, strongly depend on the
location in both longitude and latitude on the planet ($\theta$, $\phi$).
We calculate the escape rate as the sum of $\dot{M}(\theta,\phi)$,
the escape flux as a function of longitude and latitude.
Even with a uniform temperature, the
escape flux per unit area is larger toward the star and in the opposite
direction (Fig.~\ref{lobe_3d}).
 
Finally, using the above, we estimate the total escape flux of
atomic and molecular hydrogen as a function of the upper atmosphere
temperature for 
HD\,209458b (Fig.~\ref{figure2}). 
For temperature above
8\,000~K (Model A), 7\,000~K (Model B) and 6\,000~K (Model C), 
the H\,{\sc i} escape flux is larger than the minimum flux of
$10^{10}$~g$\,$s$^{-1}$ needed to explain the occultation depth
of 15\% as observed in Lyman~$\alpha$ (Vidal-Madjar et al.\ 2003).
From the escape rate, we can derive
the corresponding life time needed to evaporate the total mass of the
planet $t_1=M_{\rm p} / \dot{M}$. However when the planet mass
decreases, the evaporation rate increases. This results in a shorter life
time given by $t_2 = \int dM/\dot{M}$, which is typically shorter than
$t_1$ by a factor of 5 to 10 (Fig.~\ref{figure2}b).

\section{The temperature of the upper atmosphere}
\label{estimate Tup}

\begin{figure}
\includegraphics[width=\columnwidth]{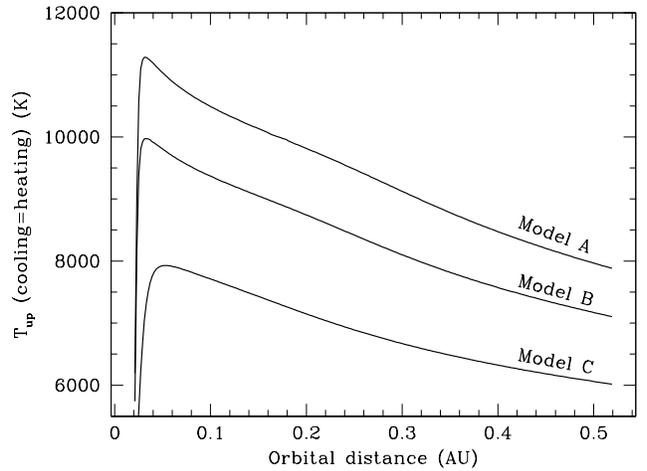}
\caption[]{Upper atmosphere temperature estimated from the energy balance
as a function of the orbital distance of a 0.69\,M$_{\rm Jup}$ planet.
The input energy is 
$2.7 \times (d/1{\rm AU})^{-2}$\,erg\,cm$^{-2}$\,s$^{-1}$.
Below 0.04\,AU, the temperature decreases because
the large escape rate cools down the thermosphere.
}
\label{figure3}
\end{figure}

Although the mechanism responsible for the heating of
the upper atmospheres in the Solar system is not fully
identified, we obtain a plausible estimate of $T_{\rm up}$ from a
comparison of heating and cooling mechanisms. A lower limit of the
heating can be estimated from the energy flux of both the stellar 
extreme ultraviolet (EUV) and Lyman\,$\alpha$ photons. Using the EUV and a
monte-carlo simulation of the multiple scattering of Lyman\,$\alpha$ photons
in the thermosphere, we estimate the minimum energy input to be 
$G = 2.7 \times (d/1\,{\rm AU})^{-2}$\,erg\,cm$^{-2}$\,s$^{-1}$, 
where $d$ is the orbital distance to the star.  
The cooling is due to a combination of the
heat conduction toward the cooler lower atmosphere, collisional exitation of
the H\,{\sc i} electronic levels, collisional ionization (photo-ionization is 
negligible), and cooling by
escaping atoms and molecules carrying off their kinetic energy.
We derive the ionization fraction and the collisonal cooling  
following Spitzer (1978) and Osterbrock
(1989) for atomic and molecular hydrogen clouds. 
With the HD\,209458b characteristics, 
the ionization fraction is barely above 0.01.
We calculated the heat conduction following the
derivation given by Hunten \& Dessler (1977). 

If we apply this calculation to Jupiter with UV heating only, we find
a temperature rise at the thermobase 
as low as $\sim$15~K (Strobel \& Smith 1973); clearly other heating mechanisms
are at work in Jupiter (Hunten \& Dessler 1977). 
Considering the UV heating only, the energy balance provides a lower limit to 
the thermospheric temperature 
(Fig.~\ref{figure3}). 
For HD\,209458b, we evaluate the lower limit of $T_{\rm up}$ 
to be about 11\,100\,K, 9\,800\,K and 7\,900\,K for Models A,
B and C, respectively. Cooling is dominated by atmospheric escape 
and collisonal excitation of H\,{\sc i}.
With these temperatures, the present evaporation rates 
of atomic and molecular hydrogen 
from HD\,209458b are estimated to be
$\dot{M_{\rm HI}} \simeq 5.7 \times 10^{10}$\,g\,s$^{-1}$ and 
$\dot{M_{\rm H_2}} \simeq 1.7 \times 10^{10}$\,g\,s$^{-1}$ in Model~A, 
$\dot{M_{\rm HI}} \simeq 1.3 \times 10^{11}$\,g\,s$^{-1}$ and 
$\dot{M_{\rm H_2}} \simeq 3.5 \times 10^{9}$\,g\,s$^{-1}$ in Model B, 
and 
$\dot{M_{\rm HI}} \simeq 5.2 \times 10^{11}$\,g\,s$^{-1}$ in Model C,
in agreement with the observational lower limit of 
$\sim$$10^{10}$\,g\,s$^{-1}$ 
(Vidal-Madjar et al.\ 2003). 
We obtain a life time ($t_2$) of $10^{10}$ to $10^{11}$\,years.
During $5 \times 10^9$ years, 
HD\,209458b may have lost 1\% to 7\% of its initial mass.

\begin{figure}
\includegraphics[height=\columnwidth,angle=90]{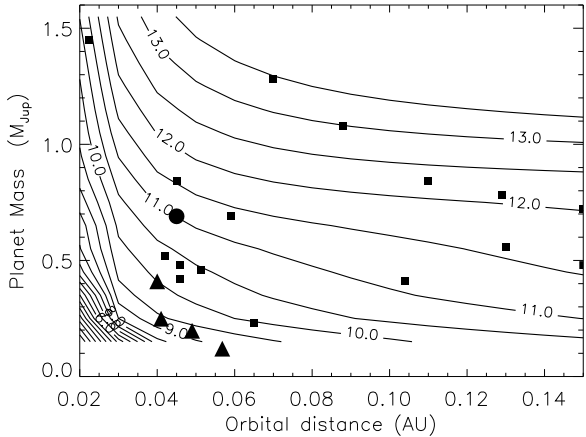}
\caption[]{Contour plot of the planet life time ($\log _{10} t_2/{\rm year}$) 
as a function of the mass and orbital distance (Model A). 
The temperature is 
obtained from the energy balance and is varying with time.
We obtain similar numerical estimates and 
contour plots for Model B and C.
The squares show the positions of detected planets. 
OGLE-TR-56b is at the upper left conner.
HD\,209458b is shown by a circle. 
HD$\,$49674b, HD$\,$46375b, HD$\,$76700b, and HD$\,$83443b (triangles)
have life time of about $10^9$~years.
These planets must have lost a large fraction of their hydrogen and might be 
remnants of former hot Jupiters. 
}
\label{plot_dm}
\end{figure}

With models of heat conduction within escaping atmospheres,
Watson et al.\ (1981) showed that the escape rate can be limited 
by heat exchange.
Using 
their result applied to HD\,209458b, we found that the escape
rate is heat-limited to $\sim$$10^{12}$\,g\,s$^{-1}$, 
in agreement
with the results obtained by Trilling (1999) and Lammer et al.\ (2003).
Although these results neglect
the tidal forces, 
our values are below this limit and therefore do not need a full calculation 
of the heat exchange within the thermosphere.

At a given temperature, the escape rate is roughly proportional
to the thermobase densities and the energy input. However,
using the temperature calculated from the energy balance,
the escape rate is less sensitive to the assumed input parameters. 
It is roughly proportional to the thermobase densities 
to the power of 0.3, and proportional to the energy input to the power
of about 0.4.

\section{Life time of hot Jupiters}

Finally, we can also estimate the life time of a given planet as a
function of its mass and orbital distance (Fig.~\ref{plot_dm}). 
We conclude that
planets with orbital distance lower than 0.03-0.04\,AU (corresponding
to orbital periods shorter than 2-3 days) have short
life time unless they are significantly heavier than Jupiter. 
This may explain why only few planets have been
detected with periods below 3 days. Low-mass hot Jupiters have
also short life times, meaning that their nature must evolve with time. 
These planets must loose a large fraction of their hydrogen.
This process can lead to planets with an hydrogen-poor atmosphere 
(``hot Neptunes''), or even with no more atmosphere at all.
The emergence of planets modified by evaporation (and possibly the
emergence of the inside core of former and evaporated hot
Jupiters) may constitute a new class of planets 
(see also Trilling et al.\ 1998). 
If they exist, these planets could be called  the ``Chthonian'' planets 
in reference to 
the Greek deities who come from hot infernal underground
(H\'ebrard et al.\ 2003).

\begin{acknowledgements}
We warmly thank Drs.\ G.~Ballester, L.~BenJaffel, 
\& C.~Parkinson 
for very fruitful discussions.
We thank Nh\^a$\!\!_{_.}$t V\~o Tr\^an$\!\!\!\grave{ }$\ \ \  for 
discussions on myths and etymology. 
\end{acknowledgements}

\end{document}